\documentclass[12pt,preprint,amsmath,amssymb,aps,prd,nofootinbib,superscriptaddress]{revtex4}
\usepackage{graphicx,epsfig}
\usepackage{amsfonts}
\usepackage[dvips]{color}

\newcommand{\ba}{\begin{array}}
\newcommand{\ea}{\end{array}}
\newcommand{\beq}{\begin{equation}}
\newcommand{\eeq}{\end{equation}}

\def\pbn3{p_b\cdot n_3}
\def\pgn3{p_g\cdot n_3}

\def\bt{\begin{table}}
\def\et{\end{table}}
\def\bc{\begin{center}}
\def\ec{\end{center}}
\def\bi{\begin{itemize}}
\def\ei{\end{itemize}}
\def\bea{\begin{eqnarray}}
\def\eea{\end{eqnarray}}
\def\beas{\begin{eqnarray*}}
\def\eeas{\end{eqnarray*}}

\def\r {\rightarrow}

\def \csqt{i}
\def \MsqRR{\sigma_{q\bar q}^{++}}
\def \MsqRL{\sigma_{q\bar q}^{+-}}
\def \MsqLR{\sigma_{q\bar q}^{-+}}
\def \MsqLL{\sigma_{q\bar q}^{--}}
\def \MsqD{\sigma_{q\bar q}^{\pm\pm}}
\def \MsqO{\sigma_{q\bar q}^{\pm\mp}}
\def \CqQs{C_{q\bar q}}
\def \Cs2{C_s}
\def \schRR{\sigma^{++}_{gg,s}}
\def \schRL{\sigma^{+-}_{gg,s}}
\def \schLR{\sigma^{-+}_{gg,s}}
\def \schLL{\sigma^{--}_{gg,s}}

\def \schD{\sigma^{\pm\pm}_{gg,s}}
\def \schO{\sigma^{\pm\mp}_{gg,s}}
\def \Ct2{C_t}
\def \tchRR{\sigma^{++}_{gg,t}}
\def \tchRL{\sigma^{+-}_{gg,t}}
\def \tchLR{\sigma^{-+}_{gg,t}}
\def \tchLL{\sigma^{--}_{gg,t}}

\def \tchD{\sigma^{\pm\pm}_{gg,t}}
\def \tchO{\sigma^{\pm\mp}_{gg,t}}
\def \Cu2{C_u}
\def \uchRR{\sigma^{++}_{gg,u}}
\def \uchRL{\sigma^{+-}_{gg,u}}
\def \uchLR{\sigma^{-+}_{gg,u}}
\def \uchLL{\sigma^{--}_{gg,u}}

\def \uchD{\sigma^{\pm\pm}_{gg,u}}
\def \uchO{\sigma^{\pm\mp}_{gg,u}}
\def \CsCt{C_{st}}
\def \stchRR{\sigma^{++}_{gg,st}}
\def \stchRL{\sigma^{+-}_{gg,st}}
\def \stchLR{\sigma^{-+}_{gg,st}}
\def \stchLL{\sigma^{--}_{gg,st}}
\def \stchD{\sigma^{\pm\pm}_{gg,st}}
\def \stchO{\sigma^{\pm\mp}_{gg,st}}
\def \CsCu{C_{su}}
\def \suchRR{\sigma^{++}_{gg,su}}
\def \suchRL{\sigma^{+-}_{gg,su}}
\def \suchLR{\sigma^{-+}_{gg,su}}
\def \suchLL{\sigma^{--}_{gg,su}}
\def \suchD{\sigma^{\pm\pm}_{gg,su}}
\def \suchO{\sigma^{\pm\mp}_{gg,su}}
\def \CtCu{C_{tu}}
\def \tuchRR{\sigma^{++}_{gg,tu}}
\def \tuchRL{\sigma^{+-}_{gg,tu}}
\def \tuchLR{\sigma^{-+}_{gg,tu}}
\def \tuchLL{\sigma^{--}_{gg,tu}}
\def \tuchD{\sigma^{\pm\pm}_{gg,tu}}
\def \tuchO{\sigma^{\pm\mp}_{gg,tu}}
\def \ct3{\cos\theta_t}
\def \st3{\sin\theta_t}
\def \csf{\cos^4\theta_t}
\def \cst{\cos^3\theta_t}
\def \cstw{\cos^2\theta_t}

\def \sstw{\sin^2\theta_t}
\def \Mt{m_t}
\def \E1{E_1}
\def \k3{k_t}

\def \Rerho2{{\rm Re}\rho}
\def \Imrho2{{\rm Im}\rho}
\def \rerho3{{\rm Re}\rho^\prime}
\def \imrho3{{\rm Im}\rho^\prime}
\def \bt{\beta_t}
\begin{document} 
\preprint{KIAS-P12074}
\vspace*{1cm}
\title{\boldmath 
Probing chromomagnetic and chromoelectric couplings of the top quark using its
polarization in pair production at hadron colliders }
\bigskip
\author{Sudhansu S. Biswal\footnote{sudhansu.biswal@gmail.com}}
\affiliation{Department of Physics, Orissa University of Agriculture and Technology, Bhubaneswar 751003, India}

\author{Saurabh D. Rindani\footnote{saurabh@prl.res.in}} 
\affiliation{Theoretical Physics Division, 
Physical Research Laboratory, 
Navrangpura, Ahmedabad 380 009, India}

\author{Pankaj Sharma\footnote{pankajs@kias.re.kr}}
\affiliation{Korea Institute for Advanced Study\\
Hoegiro 87, Dongdaemun-gu, Seoul 130-722, Korea}
\vskip 0.5 truecm 
\begin{abstract}
\vskip 0.5 truecm 
The Tevatron, where the top quark was discovered, and the currently
functional Large Hadron Collider (LHC), with copiously produced top
pairs, enable a detailed study of top-quark properties. In particular,
they can be used to test the couplings of the top quark to gauge bosons.
Several extensions of the standard model (SM) can give rise to anomalous
couplings of the top quark to gauge bosons, in particular, the gluons.
In this work we examine how top-quark polarization, which is predicted
to be negligibly small in the SM, can be used to measure chromomagnetic
and chromoelectric couplings of the top quark to gluons. We lay special
emphasis on the use of angular distributions and asymmetries of charged
leptons arising from top decay as measures of top polarization and hence
of these anomalous couplings. Sensitivities that may be reached at the
Tevatron and the LHC are obtained. 
\end{abstract}
\pacs{}

\maketitle

\section{Introduction}
The Tevatron shut down its operations last year after 8.7 fb$^{-1}$ of accumulated data. 
The first run of the Large Hadron Collider (LHC) 
with $\sqrt{s}=7$ TeV was already completed last year. 
In that run, the LHC achieved $5.1$ fb$^{-1}$ of integrated luminosity. This year it has started at 
$\sqrt{s}=8$ TeV and has been projected to collect $15$ fb$^{-1}$ of
data. After 
completing its run at $\sqrt{s}=8$ TeV, it is expected to start running at $\sqrt{s}=14$ TeV in 2014.
With the standard model (SM) cross section for 
top-pair production at $\sqrt{s}=14$~TeV
predicted to be  around 830 pb, the LHC will provide ample
opportunity to study top properties in detail. 

The top quark is the heaviest fundamental particle discovered so far with its mass $m_t=173.2\pm 0.9$ GeV \cite{top:mass}. 
Mainly for this reason, it is considered to be a strong player in the determination of the mechanism of electroweak symmetry breaking (EWSB). 
The other consequence of its large mass is that its life time is very short and decays rapidly
before any non-perturbative QCD effects can force it into a bound state. Thus, its spin information is 
preserved in terms of the differential distribution of its decay products. So by studying the 
kinematical distributions of top decay products, it is, in principle, possible to measure top polarization 
in any top production process. 

While already enough information about the top quark is available, which
shows consistency with SM expectations, future runs at the LHC will
enable more precise determination of its properties.
The most recent experimental value of the top-pair production cross section at the Tevatron by CDF with 
4.6 fb$^{-1}$ of data is $\sigma(t\bar t)=7.5\pm0.31~(\mbox{stat})\pm0.34~(\mbox{syst})$ pb \cite{cdf-cross} for $m_t=172.5$ GeV and 
is consistent with the measurements from D\O\cite{d0-cross}. These measurements are in good agreement with the 
SM prediction of $\sigma(t\bar{t})^{NNLO}_{SM} = 7.08^{+0.00+0.36}_{-0.24-0.27}$~pb for $m_t =
173$~GeV~\cite{kidonakis-tcross}. The $t\bar t$ cross section has also
been measured at the LHC, with a value of $161.9\pm
2.5$(stat)$^{+5.1}_{-5.0}$(syst) from CMS for an integrated
luminosity of 2.3 fb$^{-1}$\cite{cms-cross}, and $186\pm 13$(stat)$\pm
20$(syst)$\pm 7$(lum)
from ATLAS \cite{atlas-cross}, for an integrated luminosity of 2.05 fb$^{-1}$, in agreement
with predictions of the SM.  

There seem to be hints of new physics from the  study of top-pair production 
at the Tevatron in the forward-backward asymmetry of the top quark  
beyond the SM. Recent measurements by CDF \cite{Aaltonen:2008hc} and D\O
\cite{Abazov:2011rq} give a larger value for the asymmetry
than predicted by the SM. 

Experiments at the Tevatron and the LHC have also produced results 
on top spin 
correlations \cite{cdf-spin,d0-spin,atlas-spin,cms-spin}. 
The LHC also has results on top polarization \cite{LHC-toppol}.
These are consistent with expectations from SM. Particularly, top
polarization in the SM is predicted to be nearly vanishing at the LHC
because the dominant contributions come from strong interactions, and
are therefore parity conserving. Thus any deviation from zero would
signal physics beyond SM.
The errors are however still large, and new physics is not precluded.
In these experiments, polarization is determined by studying the decay
distribution in the rest frame of the top quark. The reconstruction of
the rest frame entails loss of accuracy. As we will see later, direct
observation of the decay distributions in the laboratory frame can be used to
probe polarization, and hence infer details of the production mechanism 
of the top. It is to be expected that this method will suffer from less
systematic uncertainties.

Top polarization and its usefulness in the study of new physics scenarios has been extensively 
treated in the literature (for some recent papers in the context of
hadron colliders, see \cite{Godbole:2010kr,Huitu:2010ad,Rindani:2011pk,
Rindani:2011gt,Baglio:2011ap,Fajfer:2012si}).
For example, in Ref. \cite{Godbole:2010kr}, it was shown how top
polarization could be utilized to 
probe the $Z^\prime$ couplings in the Little Higgs (LH) Model. In Ref.
\cite{Huitu:2010ad}, the authors showed how  
top polarization may be used to determine the parameters of the 
two Higgs Doublet Model (THDM) and minimal supersymmetric 
extension of standard model (MSSM). The effect of anomalous $Wtb$ couplings on top polarization in single-top 
production has been studied in Ref. \cite{Rindani:2011pk}. Probe of CP violation in single-top production 
using the polarization of top has been discussed in Ref. \cite{Rindani:2011gt}. 
Refs. \cite{top:fba} suggest utilizing 
top polarization as a probe of models for the top forward-backward
asymmetry observed at the Tevatron.

In this work, we study top-pair production at the Tevatron and the LHC 
in the presence of anomalous gluon couplings to a $t\bar t$ pair. In
particular, we examine the possibility of using 
top polarization and other kinematical observables constructed from
top decay products in the laboratory frame to measure these anomalous
couplings. Our main emphasis will be to show how these laboratory-frame
observables can be used to constrain the anomalous couplings. However,
since these observables arise from top polarization, they would be a
measure of top polarization as well. We therefore first discuss how
polarization can give a handle on anomalous couplings.

Top chromomagnetic and chromoelectric couplings which we study here
could arise in the SM or from new interactions at loop level.
While the CP-conserving chromomagnetic coupling can arise in the SM at
one-loop \cite{Martinez:2007qf}, the CP-violating chromoelectric 
coupling can only be generated at 3-loop level in the SM. 
Chromomagnetic and chromoelectric 
dipole moments of the top have been calculated at loop level
in various new physics models such as MSSM \cite{mssm}, THDM \cite{2hdm},
LH model \cite{LH} and in 
models with unparticles \cite{unparticle}.

We calculate our 
observables at the Tevatron and at the LHC with  centre-of-mass (cm) energies of 7 TeV (LHC7), 
8 TeV (LHC8) and 14 TeV (LHC14). We also look at the sensitivities achieved in all these scenarios 
including  statistical uncertainties with integrated luminosities of 8 fb$^{-1}$ at 
the Tevatron, 5 fb$^{-1}$ at LHC7, 10 fb$^{-1}$ at LHC8 and 10 fb$^{-1}$ for the case of LHC14. 

Anomalous $ttg$ 
couplings have been studied by several authors in the context of
top-pair \cite{atwood,haberl,hioki,saha,hioki2,hesari,gupta}, 
top-pair plus jet \cite{Cheung:1995nt} and single-top production \cite{Rizzo:1995uv} at hadron colliders. 
In Ref. \cite{Cheung:1996kc}, the author has used spin correlations in top-pair production at hadron colliders to probe 
chromomagnetic and chromoelectric dipole moments of top quarks. CP violation in top-pair production at 
hadron colliders including top chromoelectric couplings is studied in \cite{Zhou:1998wz}.

Apart from having a direct effect on top-pair production at hadron
colliders,
chromomagnetic and chromoelectric dipole couplings 
can have an indirect effect and modify the decay rate of $b\to s\gamma$ at loop level 
\cite{Hewett:1993em,b2sg}. Using the measured branching ratio
 Br($b\to s\gamma$) \cite{b2sg}, tight bounds on the chromomagnetic dipole coupling 
$\rho$ were extracted, viz., $0.03 < \rho < 0.01$.

At the Tevatron and at the LHC, the dominant process of top production, viz.,  
top-pair
production, takes place through chirality-conserving 
QCD couplings in the SM. Thus, in the SM, the 
top polarization in top-pair production can only occur through the
electroweak quark-antiquark annihilation into a virtual $Z$ and is
negligibly small.
Any new physics in which new 
couplings to top are chiral can increase top polarization. The measurement of top polarization is thus 
an important tool to study new physics in top-pair production. However, top polarization can only be measured through 
the distributions of its decay products. Hence, any new physics in top decay may contaminate the measurement of 
top polarization and, therefore of the new physics contribution in top production. Assuming only SM particles, any new physics in top decay can be 
parameterized in terms of anomalous $tbW$ couplings as
\begin{equation}
 \Gamma^\mu =\frac{-ig}{\sqrt{2}}V_{tb}\left[\gamma^{\mu}(f_{1L} P_{L}+f_{1R} P_{R})+
\frac{i \sigma^{\mu \nu}}{m_W}(p_t -p_b)_{\nu}(f_{2L} P_{L}+f_{2R} P_{R})\right] \label{anomaloustbW}
\end{equation}
where in SM $f_{1L}=1$ and $f_{1R}=f_{2L}=f_{2R}=0$.
Under the assumptions that (i) anomalous $tbW$  couplings are small, (ii) 
the top is on-shell and (iii) $t\r bW^+$ is the only decay channel, it was shown in Refs. \cite{Godbole:2006tq}
that the charged-lepton angular distributions are independent of 
the anomalous $tbW$ couplings. Thus, one can say that the charged-lepton angular distributions are clean 
and uncontaminated probes of top polarization and thus of any new physics responsible for top production.

For the above reasons, we choose, apart from top polarization,
an asymmetry constructed out of the azimuthal distribution of charged
leptons arising from top decay.

In our work, we concentrate on the leptonic decay  state arising from
either $t$ or $\bar t$ in top pair production. That is, we look at
observables constructed from the charged lepton produced in $t$ ($\bar
t$) decay, while the $\bar t$ ($t$), can decay into either a leptonic
or a hadronic final state. Often we do not distinguish between observables related to 
$t$ and those related to $\bar t$. Thus measurements made for the top
quark could also be made for the top antiquark in the process of
top-pair production, and the results combined.
However, in this case, information on CP violation would be lost. For
the specific case of measurement of the CP-violating chromoelectric
coupling, $t$ and $\bar t$ observables have to be treated separately,
and the corresponding partial cross sections appropriately added or
subtracted.

The rest of the paper is organized as follows. In the next section we
discuss the formalism and the framework of our work. In Section
\ref{production} we discuss the application of the framework to the
process of inclusive top-pair production at the Tevatron and the LHC, 
and present our results for the observables like top polarization, charged-lepton
angular distributions and the lepton azimuthal asymmetry.                      
Section \ref{sens} deals with the
statistical sensitivity of our observables to the anomalous couplings.
The following section contains the conclusions. The Appendix lists 
the production spin density matrix elements at the parton level for 
$gg$ and $q\bar q$ initial states.

\section{The framework}

We now describe the formalism underlying our analysis. 

We define the top-quark anomalous couplings to gluons including chromomagnetic and chromoelectric 
dipole form factors by the $t\bar t g$ vertex
\beq\label{coup}
\Gamma^\mu=\frac{g_s}{m_t}\ \sigma^{\mu\nu}\left(\rho+i\rho^\prime\gamma_5\right)q_\nu,
\eeq
where $\rho$ and $\rho^\prime$ are the chromomagnetic and chromoelectric 
form factors respectively, $q_\nu$ is momentum of the gluon and $m_t$ is the mass of top quark. 
Of these, the $\rho$ term is CP conserving, whereas the $\rho^\prime$ term is 
CP violating. We will treat the form factors $\rho$ and $\rho^\prime$ as
complex. Moreover, even though these form factors are in principle
energy dependent functions, we will work in the approximation that they
are constant. We will therefore often refer to them as ``couplings".
In the SM, both $\rho$ and $\rho^\prime$ are zero at tree level. 

For the calculation of the final-state charged-lepton distributions
arising from either of $t$ or $\bar t$, we use the spin density matrix formalism. 
Since the top width of about $1.5$ GeV is small compared to its mass, 
the narrow-width approximation (NWA) 
\beq\label{nwa}
\left|\frac{1}{p^2-m_t^2+im_t\Gamma_t}\right|^2\approx \frac{\pi}{m_t\Gamma}\delta(p^2-m_t^2).
\eeq
can be utilized to factor the squared amplitude into production and decay parts as
\begin{eqnarray}
\overline{|{\cal M}|^2} = \frac{\pi \delta(p_t^2-m_t^2)}{\Gamma_t m_t}
\sum_{\lambda,\lambda'} \rho^{\lambda\lambda'} \ \Gamma^{\lambda\lambda'},
\end{eqnarray}
where $\rho^{\lambda\lambda'}$ and $\Gamma^{\lambda\lambda'}$ are respectively the $2 \times 2$ top production and 
decay spin density matrices and $\lambda,\lambda' =\pm$ denote the sign of the top helicity. 
The density matrices may be defined in terms of the spin dependent
amplitudes as follows:
\begin{equation}
\rho^{\lambda\lambda'} = \sum_{\mu}M^{\rm prod}_{\lambda \mu}M^{{\rm
prod} *}_{\lambda' \mu},
\end{equation}
\begin{equation}
\Gamma^{\lambda\lambda'} = M^{\rm decay}_\lambda M^{{\rm
decay}*}_{\lambda'}
\end{equation}
Here $M^{\rm prod}_{\lambda \mu}$ is the amplitude for top pair
production, with the sign of top helicity $\lambda$, and that of the
antitop helicity $\mu$. $M^{\rm decay}_{\lambda}$ is the amplitude for
the decay of the top with helicity $\lambda$.
Analogous expressions may be written down for the density matrices for
the antitop quark.

After phase space integration of $\rho^{\lambda\lambda'}$ we get the resulting polarization density matrix 
$\sigma^{\lambda\lambda^\prime}$. The (1,1) and (2,2) diagonal elements of $\sigma^{\lambda\lambda^\prime}$ 
are the cross sections  for the production of positive and negative helicity tops and 
$\sigma_{\rm{tot}}=\sigma^{++}+\sigma^{--}$ is the total cross section.

Using Eq. (4) we can write the partial cross section in the parton cm frame as 
\bea
d\sigma&=&\frac{1}{32(2\pi)^4 \Gamma_t m_t} 
\int \left[ \sum_{\lambda,\lambda'}
\frac{d\sigma^{\lambda\lambda'}}{d\cos\theta_t} \
\left(\frac{\langle\Gamma^{\lambda\lambda'}\rangle}{p_t \cdot p_\ell}\right) \right] \ 
\nonumber \\
&\times& 
d\cos\theta_t \ d\cos\theta_\ell \ d\phi_\ell\ E_\ell dE_\ell \ dp_W^2,
\label{dsigell}
\eea
where the $b$-quark energy integral is replaced by an integral 
over the invariant mass $p_W^2$ of the $W$ boson, its polar-angle
integral is carried out using the Dirac delta function of Eq.
(\ref{nwa}), and the average over its azimuthal angle is denoted by the
angular brackets.
We obtain analytical expressions for the spin density matrix for
top-pair production including the 
contributions of anomalous $t\bar t g$ couplings to linear order at the
parton level. These expressions for $gg$ and $q\bar q$ initial states
are listed separately in the Appendix. 
Use has been made of the analytic manipulation program 
\texttt{FORM} \cite{Vermaseren:2000nd}. 
The expressions for the top-decay 
spin density matrix has been evaluated without linear approximation in anomalous $tbW$ couplings in 
Ref. \cite{Rindani:2011pk}. However, since we plan to work to linear
order also in the $tbW$ anomalous couplings, and evaluate observables
dependent only on lepton angular variables, we need not include the
dependence on $tbW$ anomalous couplings.

\section{The top-pair production process}\label{production}
We make use of the analytical expressions for the spin density matrix for $t\bar t$ production including 
anomalous $ttg$ couplings to linear order listed in the Appendix. 
QCD gauge invariance of the $ttg$ anomalous couplings requires a $ggtt$
four-point coupling, which has also been 
included in our expressions.
We find that at linear order, the real part of the coupling $\rho$ and the imaginary part of the 
coupling $\rho^\prime$ give significant contributions to the diagonal elements of  
production density matrix, which are the ones which contribute to top
polarization. The off-diagonal elements of the matrix get contributions from
Im$\rho$ and Im$\rho^{\prime}$, but not from the real parts of the
anomalous couplings. 
The parton-level distributions
are convoluted with parton distributions, which we do numerically.

We neglect all fermion masses except that of the top and set $V_{tb}=1$. 
For numerical calculations, we use the leading-order parton distribution function 
(PDF) set \texttt{CTEQ6L} \cite{cteq6}
with a factorization scale of $m_t=173.2$ GeV. We also evaluate the strong
coupling at the same scale, $\alpha_s(m_t)=0.1085$. 
We make use of the 
following values for other parameters: $M_W=80.403$ GeV, the electromagnetic 
coupling $\alpha_{em}(m_Z)=1/128$ and $\sin^2\theta_W=0.23$.
 We neglect the electroweak contributions in the production process.
We take only one coupling to be non-zero at a time in the analysis except in Section \ref{sens} 
where we show how simultaneous limits on two of anomalous $ttg$ couplings may be obtained. 
In evaluating the angular distribution of the charged lepton from top
decay, we impose the 
acceptance cuts $p_T^\ell>20$ GeV and $|\eta|<2.5$ on the transverse
momentum $p_T^\ell$ and rapidity $\eta$ of the charged lepton.

\subsection{Top polarization}

The degree of longitudinal polarization $P_t$ of the top quark is given by
\begin{equation}
P_t=\frac{\sigma^{++}-\sigma^{--}}{\sigma^{++}+\sigma^{--}},
\label{eta3def}
\end{equation}
with an analogous expression for the polarization of the top antiquark.
\begin{figure}[h]
\begin{center}
\includegraphics[angle=270,width=4.5in]{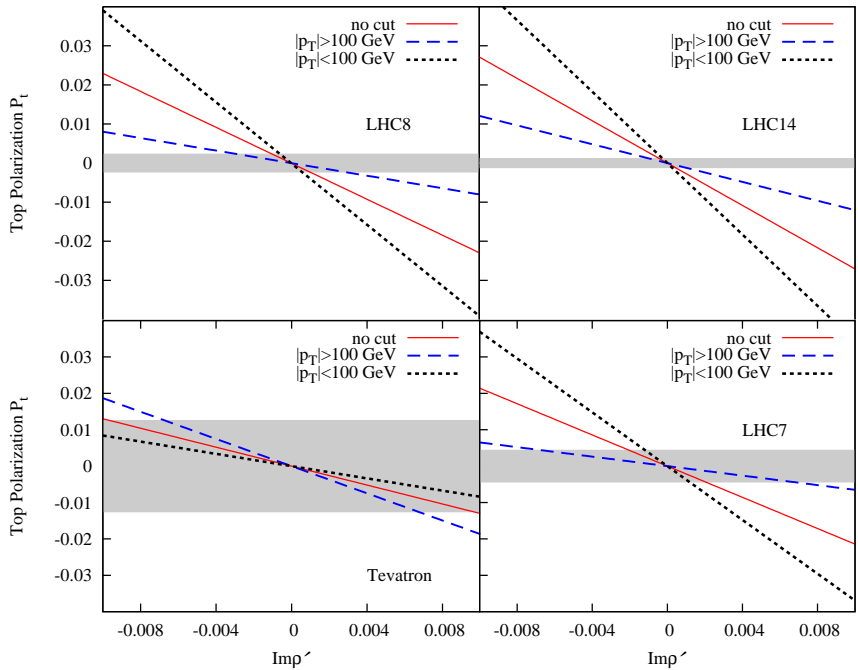} 
\caption{ The top polarization $P_t$ in $t\bar t$  production at the Tevatron (bottom left), LHC7 (bottom right),
LHC8 (top left) and LHC14 (top right) as a function of the anomalous $ttg$ coupling Im$\rho^\prime$. 
The grey band shows the 3$\sigma$ error interval in the SM without any $p_T$ cut.} 
\label{polcoup}
\end{center}
\end{figure}

In the SM, $P_t$ is 
predicted to be zero at tree level for top-pair production neglecting the contributions of 
s-channel $\gamma,~Z$ exchange in $q\bar q$ annihilation. We find that 
including non-vanishing anomalous $ttg$ couplings, there can be non-zero 
top polarization asymmetry. In the expressions for the spin density matrix for top-pair 
production, we see that the contributions of Im$\rho^\prime$ have opposite signs in (1,1) and (2,2) elements and hence 
leads to non-zero $P_t$ while the contributions of Re$\rho$ have the same sign in these elements and thus do not 
contribute to top polarization. Thus top polarization can be utilized to measure the coupling Im$\rho^\prime$ 
independently of all other anomalous $ttg$ couplings. The diagonal
elements of the density matrix
for $\bar t$ also show that the polarization of $\bar t$ is the same as
that of $t$, confirming that $P_t + P_{\bar t}$ is indeed a measure of
CP violation, proportional to the CP-odd coupling Im$\rho^\prime$.
$P_t$ is shown in Fig. \ref{polcoup} as a function of 
anomalous coupling Im$\rho^\prime$ in the linear approximation, for the Tevatron 
and for the LHC7, LHC8 and LHC14. 
The grey bands in the figures denote the 3$\sigma$ statistical uncertainty in 
the measurement of $P_t$. The grey band is the thinnest for the 14 TeV LHC because of 
the largest cross section and therefore the smallest statistical error.

We also study the effect of top-$p_T$ cut on top polarization. In Fig. \ref{polcoup}, we show top 
polarization for two different values of $p_T$ cut i.e. $p_T>$ 100 GeV and $p_T<$ 100 GeV. We find that for 
low-$p_T$ tops, the top polarization is larger compared to high-$p_T$ tops for the LHC while for the Tevatron,
this observation is opposite. At Tevatron, high-$p_T$ tops tend to have higher degree of polarization. 

We can understand the observation regarding the Tevatron as follows: 
At the Tevatron, the $q\bar q$  contribution dominates. In the
diagonal elements of the spin density matrix for the $q \bar
q$-initiated contribution shown in the Appendix, the coefficient of
$\imrho3$ is proportional to $\sin^2\theta_t = (p^t_T/p^t)^2$. It is
the $\imrho3$ which gives rise to the polarization, and so $P_t$ is
proportional to $(p^t_T)^2$ at the Tevatron, and it increases with
transverse momentum. As for the LHC, the result is not so easy to see.

\subsection{Angular distributions of the charged lepton}
Top polarization can be determined through the angular distribution of its decay products. In the SM, the dominant decay mode is $t\to b W^+$, 
with a branching ratio (BR) of 0.998, with the $W^+$ subsequently decaying to $\ell^+ \nu_\ell$ (semileptonic decay, BR 1/9 for each lepton) or 
$u \bar{d}$, $c\bar{s}$ (hadronic decay, BR 2/3). The angular distribution of a decay product $f$ for a top-quark ensemble has the form 
 \begin{equation}
 \frac{1}{\Gamma_f}\frac{\textrm{d}\Gamma_f}{\textrm{d} \cos \theta _f}=\frac{1}{2}(1+\kappa _f P_t \cos \theta _f).
 \label{topdecaywidth}
 \end{equation}
 Here $\theta_f$ is the angle between fermion $f$ and the top spin vector in the top rest frame and $P_t$ (defined in Eq. (\ref{eta3def})) 
is the degree of polarization of the top-quark ensemble. $\Gamma_f$ is the partial decay width and $\kappa_f$ is the spin analyzing power of $f$. 
Obviously, a larger $\kappa_f$ makes $f$ a more sensitive probe of the top spin. The charged lepton and the $d$ quark are the best spin analyzers 
with $\kappa_{\ell^+}=\kappa_{\bar{d}}=1$, while $\kappa_{\nu_\ell}=\kappa_{u}=-0.30$ and $\kappa_{b}=-\kappa_{W^+}=-0.39$, all $\kappa$ 
values being at tree level \cite{Bernreuther:2008ju}. Thus the $\ell^+$ or $d$ have the largest probability of being emitted in the 
direction of the top spin and the least probability in the direction opposite to the spin. Since at the LHC, the lepton energy and 
momentum can be measured with high precision, we focus on leptonic decays of the top. 

To reconstruct the top-rest frame, one needs full 
information about top momentum. However, due to the missing neutrino, it
is not possible to  reconstruct  completely and unambiguously 
the top longitudinal momentum and thus, this incomplete information may lead to large systematic errors. In this work, we focus 
on laboratory-frame angular distributions of the charged lepton and thus do not require full determination of the top momentum. In this sense, 
the observables we construct are more robust against systematic errors.
Also, as mentioned earlier and shown in Refs. \cite{Godbole:2006tq}, the charged-lepton angular distribution in the lab frame is independent of 
any new physics in top decay and  is thus a clean and uncontaminated probe of new physics in top production.

We first obtain the angular distribution of the charged 
lepton in the parton cm frame, by integrating over the lepton energy, with
limits given by $m_W^2<2(p_t\cdot p_\ell) < m_t^2$. This integral can be done analytically, giving the following expression for 
the differential cross section in the parton cm frame:
\bea\label{angdist}
\frac{d\sigma}{d\cos\theta_t \ d\cos\theta_\ell \ d\phi_\ell} &=&  \frac{1}{32 \ \Gamma_t m_t} \ \frac{1}
{(2\pi)^4}\int \left[ \sum_{\lambda,\lambda'} \frac{d\sigma^{\lambda\lambda'}}{d \cos\theta_t} 
 g^4 \mathcal{A}^{\lambda\lambda'} \right]
|\Delta(p_W^2)|^2 dp_W^2,
\eea
where 
\bea\label{angmat1}
 \mathcal A^{\pm\pm}&=&\frac{m_t^6}{24(1-\beta_t
\cos\theta_{t\ell})^3 E_t^2}
\Big[(1-r^2)^2  (1\pm\cos\theta_{t\ell})(1\mp\beta_t)(1+2r^2)\Big],\\
\label{angmat2}
\mathcal A^{\pm\mp}&=&\frac{m_t^7}{24(1-\beta_t
\cos\theta_{t\ell})^3 E_t^3}\sin\theta_{t\ell} e^{\pm i\phi_\ell}
\Big[(1-r^2)^2(1+2r^2)\Big].
 \eea
Here $r=m_W/m_t$ and $\cos\theta_{t\ell}$ is the angle between the top
quark and the charged lepton in top decay in the parton cm frame, 
given by
\begin{equation}
 \cos \theta_{t \ell}=\cos \theta_t \cos \theta_{\ell}+\sin \theta_t \sin \theta_{\ell} \cos \phi_{\ell},
\label{costhetatl}
\end{equation}
where $\theta_\ell$ and $\phi_\ell$ are the lepton polar and azimuthal angles.

\begin{figure}[t]
\begin{center}
\includegraphics[angle=270,width=6.in]{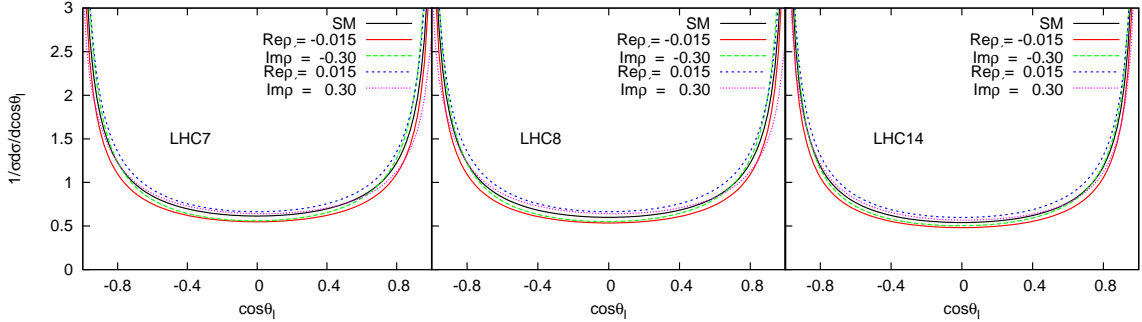} 
\caption{ The normalized polar-angle distribution of the charged lepton in $t\bar t$ 
production at the LHC7 (left), LHC8 (centre) 
and LHC14 (right) for the SM and with anomalous $ttg$ couplings.} 
\label{dist-polar}
\end{center}
\end{figure}

\begin{figure}[t]
\begin{center}
\includegraphics[angle=270,width=3.in]{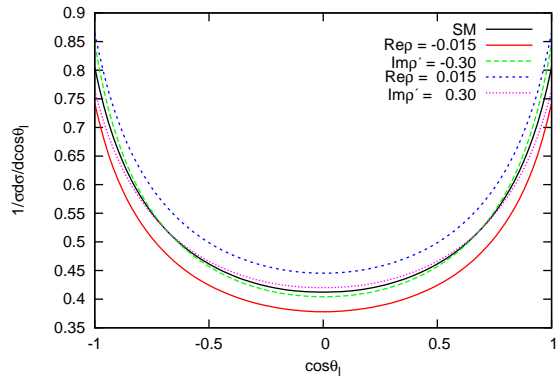} 
\caption{ The normalized polar-angle distribution of the charged lepton in $t\bar t$ 
production at the Tevatron for the SM and with anomalous $ttg$ couplings.} 
\label{dist-polar-TT}
\end{center}
\end{figure}
In the lab frame, we define the lepton polar angle w.r.t. either 
beam direction as the $z$ axis and the azimuthal w.r.t. the top-production 
plane chosen as the $x$-$z$ plane, with the 
convention that the $x$ component of the top momentum is positive. 
At the LHC, which is a symmetric collider, it is not possible to define
a positive sense for the $z$ axis. Hence lepton angular distribution is symmetric under interchange of
$\theta_{\ell}$ and $\pi-\theta_{\ell}$ as well as of $\phi_{\ell}$ and $2\pi-\phi_{\ell}$.

We first look at the polar-angle distribution of the charged lepton and  the effect on it of anomalous $ttg$ 
couplings. As can be seen from Fig. \ref{dist-polar}, where we plot the polar-angle distribution 
for LHC7, LHC8 and LHC14, the normalized distributions (here and later,
we normalize distributions to the SM cross sections) are 
insensitive to anomalous $ttg$ couplings. On the other hand, for the Tevatron, the polar-angle distribution  
are found to be somewhat sensitive as can be seen from Fig. \ref{dist-polar-TT}. The sensitivity of polar-angle 
distributions on the anomalous $ttg$ couplings have been studied in detail in Ref. \cite{hioki} for the 
Tevatron, LHC7 and LHC14. Our results for these distributions agree with them. It is interesting to note that 
even though it is possible in principle to have a forward-backward asymmetric distribution at the Tevatron, the 
chromomagnetic and chromoelectric couplings in Eq. (\ref{coup}) do not generate an asymmetry.

\begin{figure}[h]
\begin{center}
\includegraphics[angle=270,width=6.2in]{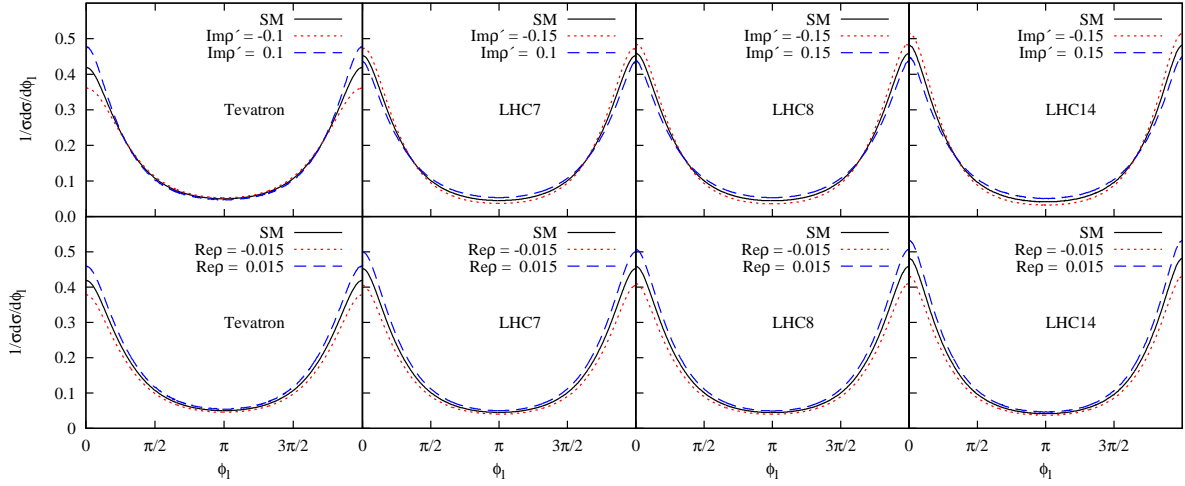} 
\caption{ The normalized azimuthal distribution of the charged lepton in $t\bar t$ 
production at the Tevatron, LHC7, LHC8 and LHC14 for anomalous $ttg$ coupling Re$\rho$ (bottom) 
and Im$\rho^\prime$. Also shown in each case is the distribution for the
SM} 
\label{dist-azi}
\end{center}
\end{figure}

We next look at the contributions of anomalous couplings to 
the azimuthal distribution of the charged lepton. In Figs. \ref{dist-azi}
we show the normalized azimuthal distribution of the charged lepton in a linear 
approximation of the couplings for Tevatron, LHC7, LHC8 and LHC14 for different values of 
Re$\rho$ and Im$\rho^\prime$ taken non-zero one at a time. 
We see that the curves for the couplings Re$\rho$ and Im$\rho^\prime$ peak near
$\phi_{\ell}=0$ and $\phi_{\ell}=2\pi$.

In principle, it is possible to separate the dependence on the two
couplings by taking the sum and difference of the distributions for $t$
and $\bar t$. The difference would be CP odd, and hence dependent only
on Im$\rho^\prime$, whereas the sum would be CP even, depending only on
Re$\rho$.


We now discuss two angular asymmetries which would serve as a measure of
the anomalous couplings. The first depends on the polar-angle distributions of
the charged leptons from $t$ and $\bar t$, and the second one on the
azimuthal distributions.

\subsection{Charge Asymmetry}
We first look at a CP-violating asymmetry which is generated by 
the difference in the
charged-lepton polar-angle distributions arising from the top and the antitop.
We define the charge asymmetry 
\begin{equation}\label{ch-asy}
A_{\rm ch}(\theta_0) = \frac{1}{2\sigma_{\bf SM} (\theta_0)}\displaystyle
\int^{\cos\theta_0}_{-\cos\theta_0}
 d\cos\theta \left( \frac{d\sigma^+}{d\cos\theta} -
\frac{d\sigma^-}{d\cos\theta} \right),
\end{equation}
where $d\sigma^{\pm}/d\cos\theta$ denote the differential cross sections for 
$\ell^+$ and $\ell^-$ production from $t$ and $\bar t$ decay
respectively, and $\sigma_{\bf SM} (\theta_0)$ is the cross section for
either $\ell^+$ or $\ell^-$ production, with a cut-off of $\theta_0$ in
the forward and backward directions of the lepton.
It is obvious that for $\theta_0=0$, the numerator of Eq. (\ref{ch-asy})
vanishes, because it measures the difference in the $\ell^+$ and $\ell^-$
production rates at all angles, which is zero from charge conservation.
However, with a cut-off $\theta_0$, $A_{\rm ch}(\theta_0)$ can be
non-zero, and is a measure of CP violation. 
It can be seen from the equations in the Appendix that $A_{\rm
ch}(\theta_0)$ is proportional to Im$\rho^\prime$.

\begin{figure}[h]
\begin{center}
\includegraphics[angle=270,width=6.5in]{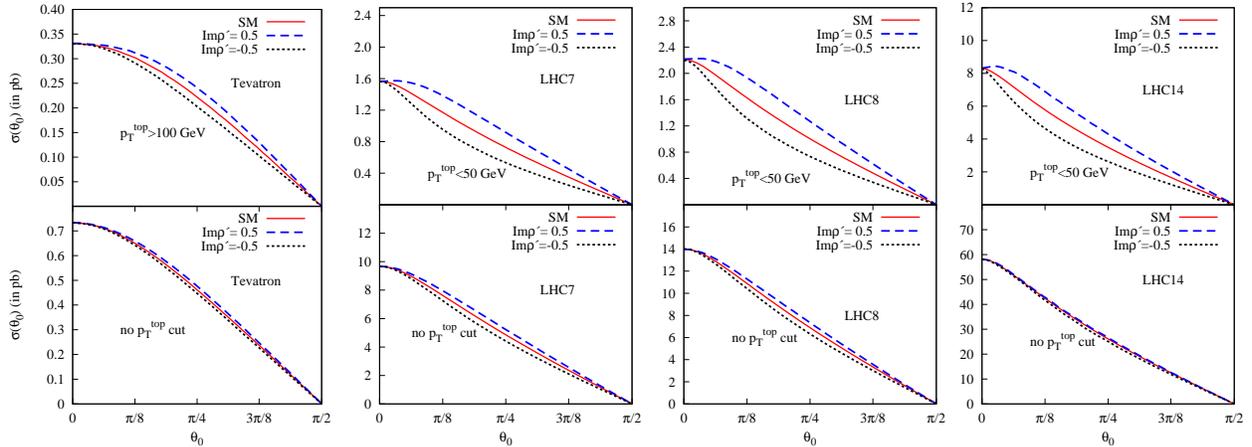} 
\caption{ The cross section as a function of cut-off angle $\theta_0$ of
the charged lepton  
in top-pair production at the Tevatron (bottom-let), LHC7 (bottom-right), LHC8 (top-left) 
and LHC14 (top-right) for anomalous $ttg$ coupling Im$\rho^\prime$. The
SM cross section is also shown in each case.} 
\label{crx-cut}
\end{center}
\end{figure}

\begin{figure}[h]
\begin{center}
\includegraphics[angle=270,width=6.50in]{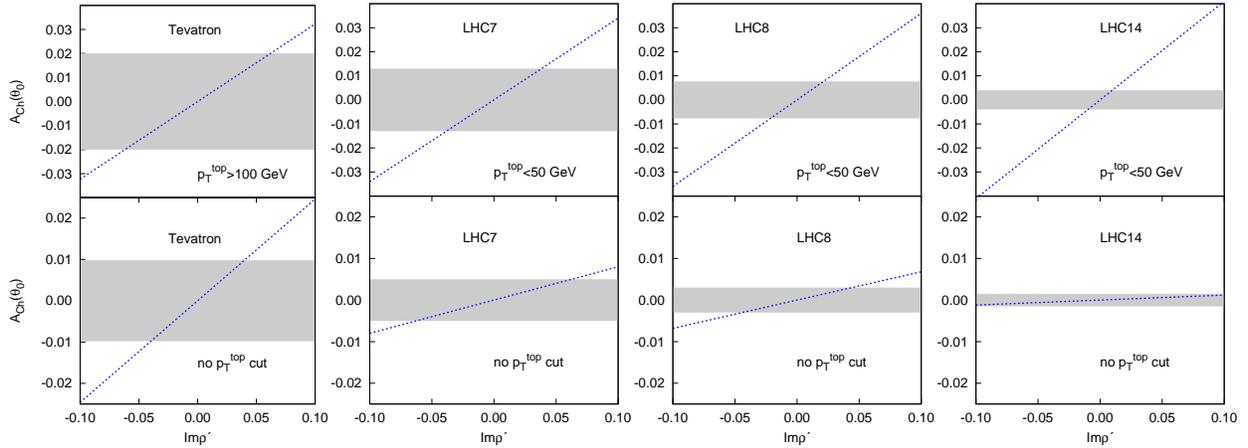}
\caption{ The charge asymmetry $A_{\rm ch}(\theta_0)$ as a function of charged lepton  
in top-pair production at the Tevatron (bottom-left), LHC7 (bottom-right), LHC8 (top-left) 
and LHC14 (top-right) for anomalous $ttg$ coupling Im$\rho^\prime$ for $\theta_0=\pi/8$.} 
\label{chAsym}
\end{center}
\end{figure}

We plot in Fig. \ref{crx-cut} the cross sections for charged leptons $\ell^\pm$ 
coming from decay of top/anti-top in top pair production as a function of cut-off 
angle $\theta_0$. We see from the Fig. \ref{crx-cut} that the deviation in the 
cross section is relatively large in the range $[\pi/8,3\pi/8]$. 
To optimize the charge asymmetry of lepton, we choose the cut-off angle
$\theta_0$ 
to be $\pi/8$ and evaluate the asymmetry as a function of Im$\rho^\prime$. 
In Fig. \ref{chAsym}, we plot the charge asymmetry of the lepton as defined in Eq. (\ref{ch-asy}) as a function of
Im$\rho^\prime$ for chosen value  $\pi/8$ of $\theta_0$
for Tevatron, LHC7, LHC8 and LHC14.

We also study the effect of top-$p_T$ cuts on the lepton charge asymmetry $A_{\rm SM}(\theta_0)$. 
From the top panel of Fig. \ref{crx-cut}, we see that at the LHC, keeping low $p_T$ top/anti-top would help 
in enhancing $A_{\rm SM}(\theta_0)$ while at the Tevatron, the reverse is true. So, we put a cut on 
top/anti-top $p_T<50$ GeV at the LHC and $p_T>100$ GeV at the Tevatron. We show the effects of 
these $p_T$ cuts on charge asymmetry in top panel of Fig. \ref{chAsym}. We find that though the 
statistical uncertainties increase due to the reduction in number of events, the asymmetry is increased 
enough times to compensate the reduction in events and thus results in the enhancement of the limits 
obtained by $A_{\rm SM}(\theta_0)$ on $\imrho3$.

\subsection{Azimuthal Asymmetry}
\begin{figure}[h]
\begin{center}
\includegraphics[angle=270,width=6.0in]{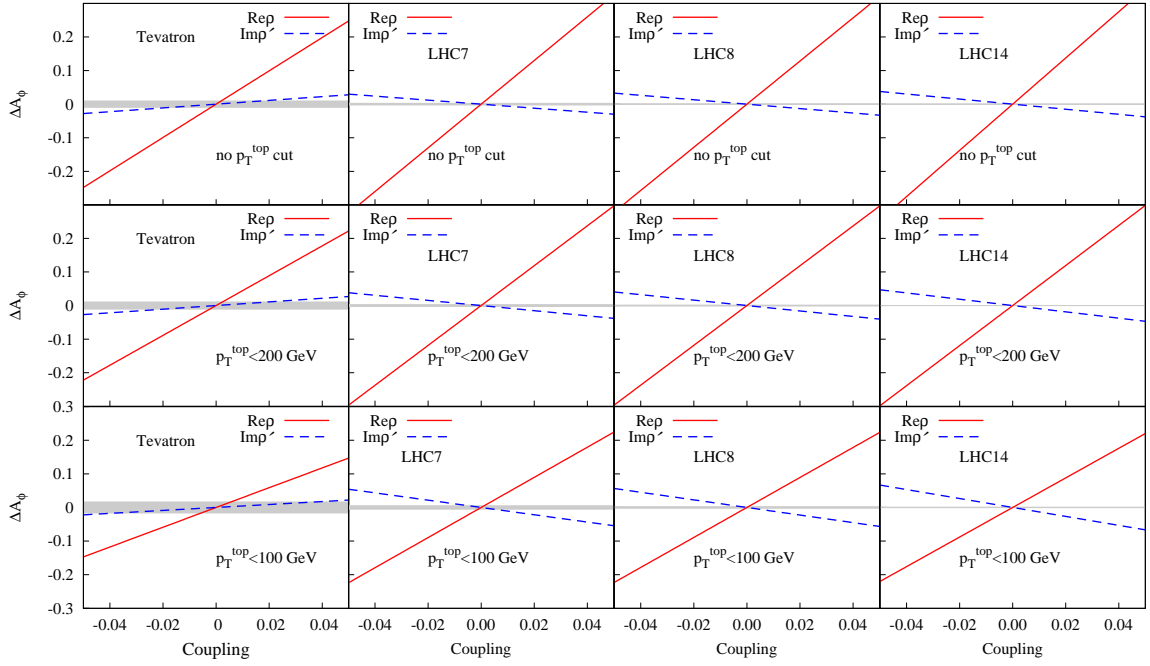} 
\caption{ The azimuthal asymmetry of the charged lepton in $t\bar t$ 
production at the Tevatron (1st column), LHC7  (2nd column), LHC8 (3rd column) 
and LHC14  (4th column) for different anomalous $ttg$ couplings.} 
\label{phiasym-lower}
\end{center}
\end{figure}

\begin{figure}[h]
\begin{center}
\includegraphics[angle=270,width=6.0in]{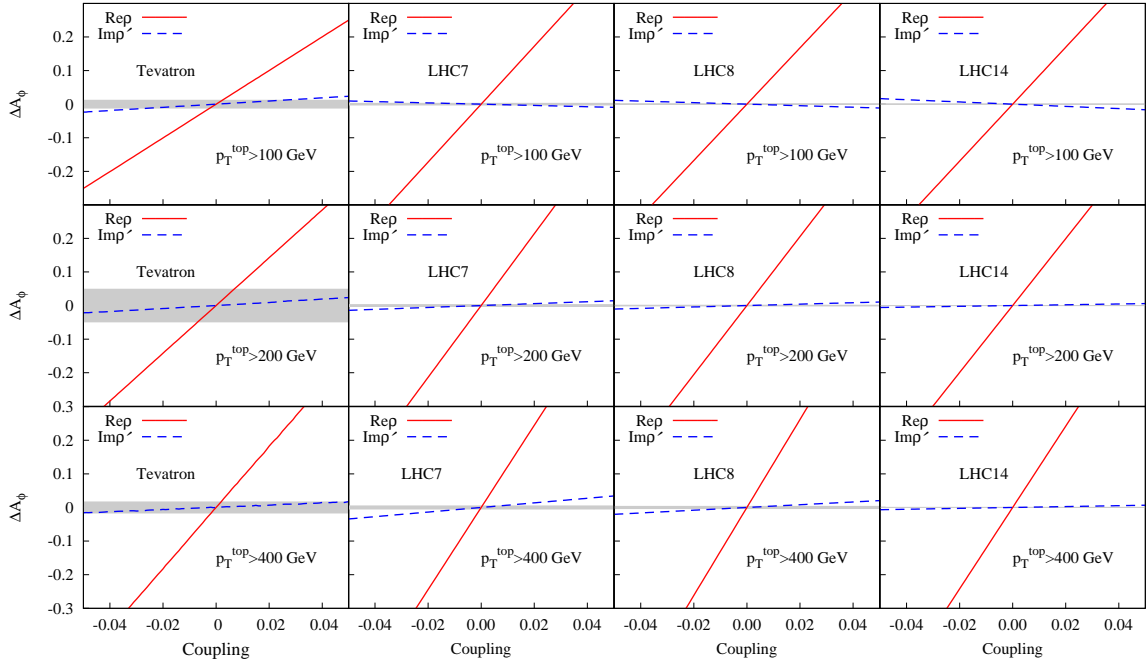} 
\caption{ The azimuthal asymmetry of the charged lepton in $t\bar t$ 
production at the Tevatron (bottom-left), LHC7 (bottom-right), LHC8 (top-left) 
and LHC14 (top-right) for different anomalous $ttg$ couplings.} 
\label{phiasym-upper}
\end{center}
\end{figure}

As can be seen from Fig. \ref{dist-azi}, the curves are well 
separated at the peaks for the chosen values of the anomalous $ttg$ couplings 
and are also well separated from the curve for the SM. We define an azimuthal asymmetry for the lepton to quantify 
these differences in the distributions by
\begin{equation}
 A_{\phi}=\frac{\sigma(\cos \phi_\ell >0)-\sigma(\cos
\phi_\ell<0)}{\sigma(\cos \phi_\ell >0)+\sigma(\cos \phi_\ell<0)},
\label{aziasy}
\end{equation}
where the denominator is the total cross section. 
This azimuthal asymmetry is in fact the ``left-right asymmetry'' of the 
charged lepton at the LHC defined with respect to the beam direction, 
with the right hemisphere defined as that in
which the top momentum lies, and the left one being the opposite one.
Plots of $A_{\phi}$
as a function of the couplings are shown in Figs. \ref{phiasym-lower} and \ref{phiasym-upper} 	 
for Tevatron, LHC7, LHC8 and LHC14. 

From Fig. \ref{dist-azi}, we see that the azimuthal distribution of the decay charged lepton 
is more sensitive to Re$\rho$ than to Im$\rho^\prime$. 
Hence, we expect that the azimuthal asymmetry we construct in Eq. (\ref{aziasy}) would be 
a sensitive probe of Re$\rho$. This fact can indeed be seen from  
Figs. \ref{phiasym-lower} and \ref{phiasym-upper} where the straight line 
for Re$\rho$ is steeper than for Im$\rho^\prime$ implying 
a more significant contribution from the former. The reason we get straight
lines for individual contributions to the asymmetry is that we are
working in a linear approximation for the anomalous couplings. 

As mentioned earlier in the context of distributions, the dependence on
the two couplings can be separated by choosing the sum and difference of
the azimuthal asymmetries for $t$ and $\bar t$. The difference being CP
odd, would be dependent only on Im$\rho^\prime$.

We also study the behavior of $A_{\phi}$ in the presence of cuts on the top transverse momentum. 
In the top panel of the Fig. \ref{phiasym-lower}, we show the behavior of $A_{\phi}$ as  
functions of Re$\rho$ and Im$\rho^\prime$ with no cut on the top transverse
momentum. In the middle 
and lower panel, we show $A_\phi$ when we consider tops with $p_T<$ 100 GeV and 200 GeV 
respectively. Similarly in Fig. \ref{phiasym-upper} we consider high-$p_T$ tops to evaluate the 
asymmetry. In the top, the middle  and the lower panel of Fig. \ref{phiasym-upper}, we show 
$A_\phi$ as functions of Re$\rho$ and Im$\rho^\prime$ for top quarks with $p_T>$ 100 GeV, 
200 GeV and 400 GeV respectively.

We find that for high-$p_T$ tops, the azimuthal distribution is relatively more peaked than for low-$p_T$ ones. 
The reason for this is the $(1-\beta_t\cos\theta_{t\ell})^3$ factor in the denominator of Eqs. (\ref{angmat1}) 
and (\ref{angmat2}). Thus, when $\beta_t$ is large, the distribution tends to peak near 0 and 2$\pi$. As a result 
asymmetry $A_\phi$ is larger for high-$p_T$ tops. This effect can be seen from Figs. \ref{phiasym-lower} and \ref{phiasym-upper} 
where it can be easily seen that as the $p_T$ of the top is increased, the azimuthal asymmetry is larger. Hence we conclude 
that asymmetry constructed from high-$p_T$  tops would be more useful in constraining the anomalous top-gluon couplings. 
From the Fig. \ref{phiasym-upper} we see that the coupling Re$\rho$ is
more sensitive at the Tevatron and LHC14. 
Though the value of the asymmetry also increases for lower-$p_T$ tops, the statistics is very low in that region 
and thus we do not gain in sensitivity.

\section{\boldmath Sensitivity analysis for anomalous $ttg$ couplings}\label{sens}
We now study the statistical significance of the observables discussed in the 
previous sections to the anomalous $ttg$ couplings at the Tevatron, LHC7, LHC8 and LHC14. 
For Tevatron, LHC7, LHC8 and LHC14, we assume integrated luminosities of 8 fb$^{-1}$, 5 fb$^{-1}$, 10 fb$^{-1}$, and 10
fb$^{-1}$ respectively. To obtain the 3$\sigma$ limit on the anomalous 
$ttg$ couplings from a measurement of an observable, we find those
values of the couplings for which the observable deviates by 3$\sigma$ from its SM value. 
The statistical uncertainty $\sigma_i$ in the measurement of any generic asymmetry $\mathcal A_i$ is given by 
\beq\label{stat-dev}
\sigma_i=\sqrt{\frac{1-\left({\mathcal A}^{SM}_i\right)^2}{\mathcal N}},
\eeq
where ${\mathcal A}^{SM}_i$ is the asymmetry predicted in the SM and $\mathcal N$ is the total number of events  
predicted in the SM. We apply this to the various asymmetries we have discussed. 
In case of the top polarization asymmetry, the limits are obtained on the assumption that the polarization can be measured with 100\% accuracy.
When lepton angular variables are used, their intrinsic efficiency to measure 
top polarization is already built in our formalism. We do not take
into account cuts which may be needed for reducing background events.
This may result in some loss of efficiency, which we have not attempted
to estimate. 

\begin{table}[h]
 \begin{tabular}{c|c|c|c|c}
  \hline
&\multicolumn{1}{c|}{$P_t$}&\multicolumn{2}{c|}{$A_\phi$}& $A_{\rm ch}(\theta_0=\pi/8)$\\
\hline
\hline
		&  Im$\rho^\prime$ 			& 	Re$\rho$	&  Im$\rho^\prime$ & Im$\rho^\prime$ \\
 \hline
 Tevatron 	& [$-9.75$, $9.75]\times$10$^{-3}$	& [$-2.22$, $2.22]\times$10$^{-2}$	& [$-1.96$, $1.96]\times$10$^{-2}$ & $[-3.98,3.98]\times10^{-2}$\\
 LHC7 		& [$-2.10$, $2.10]\times$10$^{-3}$	& [$-1.43$, $1.43]\times$10$^{-3}$	& [$-6.52$, $6.52]\times$10$^{-3}$ & $[-6.25,6.25]\times10^{-2}$\\
 LHC8 		& [$-1.06$, $1.06]\times$10$^{-3}$	& [$-3.58$, $3.58]\times$10$^{-4}$	& [$-3.50$, $3.50]\times$10$^{-3}$ & $[-4.41,4.41]\times10^{-2}$\\
 LHC14		& [$-5.59$, $5.59]\times$10$^{-4}$	& [$-1.60$, $1.60]\times$10$^{-4}$	& [$-1.46$, $1.46]\times$10$^{-3}$ & $[-1.25,1.25]\times10^{-1}$\\
 \hline
\hline
  \end{tabular}
\caption{Individual limits on anomalous couplings Re$\rho$ and Im$\rho^\prime$ which may be obtained by the measurement of the observables 
at Tevatron, LHC7, LHC8 and LHC14 with integrated luminosities of 8 fb$^{-1}$, 5 fb$^{-1}$, 10 fb$^{-1}$, and 10
fb$^{-1}$ respectively.}
 \label{lim}
 \end{table}

The 3$\sigma$ limits on Re$\rho$ and Im$\rho^\prime$ are given in Table \ref{lim} where we assume only 
one anomalous coupling to be non-zero at a time. In case of the lepton distributions, we take into account only one leptonic channel.
Including other leptonic decays of the top would improve the limits further. 

\begin{table}[h]
 \begin{tabular}{c|c|c|c|c}
  \hline
&\multicolumn{1}{c|}{$P_t$}&\multicolumn{2}{c|}{$A_\phi$}& $A_{\rm ch}(\theta_0=\pi/8)$\\
\hline
\hline
		&  Im$\rho^\prime$ 			& 	Re$\rho$	&  Im$\rho^\prime$ & Im$\rho^\prime$ \\
 \hline
 Tevatron 	& [$-1.50$, $1.50]\times$10$^{-2}$	& [$-6.12$, $6.12]\times$10$^{-3}$	& [$-4.09$, $4.09]\times$10$^{-2}$  & $-$\\
 LHC7 		& [$-1.22$, $1.22]\times$10$^{-3}$	& [$-1.45$, $1.45]\times$10$^{-3}$	& [$-5.98$, $5.98]\times$10$^{-3}$  & $[-3.82,3.82]\times10^{-2}$\\
 LHC8 		& [$-6.22$, $6.22]\times$10$^{-4}$	& [$-8.97$, $8.97]\times$10$^{-4}$	& [$-3.55$, $3.55]\times$10$^{-3}$  & $[-2.14,2.14]\times10^{-2}$\\
 LHC14		& [$-2.85$, $2.85]\times$10$^{-4}$	& [$-4.30$, $4.30]\times$10$^{-4}$	& [$-1.43$, $1.43]\times$10$^{-3}$  & $[-9.76,9.76]\times10^{-3}$\\
 \hline
\hline
  \end{tabular}
\caption{Individual limits on anomalous couplings Re$\rho$ and
Im$\rho^\prime$, with a cut $p_T<$100 GeV on the top transverse momentum  
(for $A_{\rm ch}$, we take $p_T<$50 GeV), which may be obtained by the measurement of the observables 
at Tevatron, LHC7, LHC8 and LHC14 with integrated luminosities of 8 fb$^{-1}$, 5 fb$^{-1}$, 10 fb$^{-1}$, and 10
fb$^{-1}$ respectively.}
 \label{lim-cut1}
 \end{table}

In Table \ref{lim-cut1}, we give the 3$\sigma$ limits on Re$\rho$ and
Im$\rho^\prime$ applying a cut $p_T<100$ GeV on the top transverse momentum. 
From the table, we find that though the asymmetry increases for Im$\rho^\prime$ with the cut, the limits on it do not change much 
because of the opposite effect of reduction in statistics. On the other hand, the limits on Re$\rho$ actually worsen because 
the top-$p_T$ cut reduces the asymmetry for Re$\rho$.

\begin{table}[h]
 \begin{tabular}{c|c|c|c|c}
  \hline
&\multicolumn{1}{c|}{$P_t$}&\multicolumn{2}{c|}{$A_\phi$}& $A_{\rm ch}(\theta_0=\pi/8)$\\
\hline
\hline
		&  Im$\rho^\prime$ 			& 	Re$\rho$	&  Im$\rho^\prime$ &  Im$\rho^\prime$\\
 \hline
 Tevatron 	& [$-6.79$, $6.79]\times$10$^{-3}$	& [$-1.22$, $1.22]\times$10$^{-3}$	& [$-1.87$, $1.87]\times$10$^{-2}$ & [$-6.19$, $6.19]\times$10$^{-2}$\\
 LHC7 		& [$-6.90$, $6.90]\times$10$^{-3}$	& [$-4.64$, $4.64]\times$10$^{-4}$	& [$-4.44$, $4.44]\times$10$^{-2}$ & $-$\\
 LHC8 		& [$-1.94$, $1.94]\times$10$^{-3}$	& [$-2.86$, $2.86]\times$10$^{-4}$	& [$-1.05$, $1.05]\times$10$^{-2}$ & $-$\\
 LHC14		& [$-1.08$, $1.08]\times$10$^{-3}$	& [$-1.30$, $1.30]\times$10$^{-4}$	& [$-3.33$, $3.33]\times$10$^{-3}$ & $-$\\
 \hline
\hline
  \end{tabular}
\caption{Individual limits on anomalous couplings Re$\rho$ and
Im$\rho^\prime$, with a cut $p_T>$100 GeV on the top transverse momentum, which may be obtained by the measurement of the observables 
at Tevatron, LHC7, LHC8 and LHC14 with integrated luminosities of 8 fb$^{-1}$, 5 fb$^{-1}$, 10 fb$^{-1}$, and 10
fb$^{-1}$ respectively.}
 \label{lim-cut2}
 \end{table}
In Table \ref{lim-cut2}, we give the 3$\sigma$ limits on Re$\rho$ and Im$\rho^\prime$ applying a cut $p_T>100$ GeV. 
From the table, we find that with this cut, the limits are more stringent for Re$\rho$ since the asymmetry $A_\phi$ for it is 
steeper as compared to the value without cuts. On the other hand, the limits on Im$\rho^\prime$ actually worsen because 
the top-$p_T$ cut reduces the asymmetry for Re$\rho$.

We also obtain simultaneous limits (taking both Re$\rho$ and Im$\rho^\prime$ non-zero simultaneously) 
on these anomalous couplings that may be obtained by combining the measurements at Tevatron with LHC7, LHC8 and LHC14 . 
\begin{figure}[h]
\begin{center}
 \includegraphics[width=2.0in]{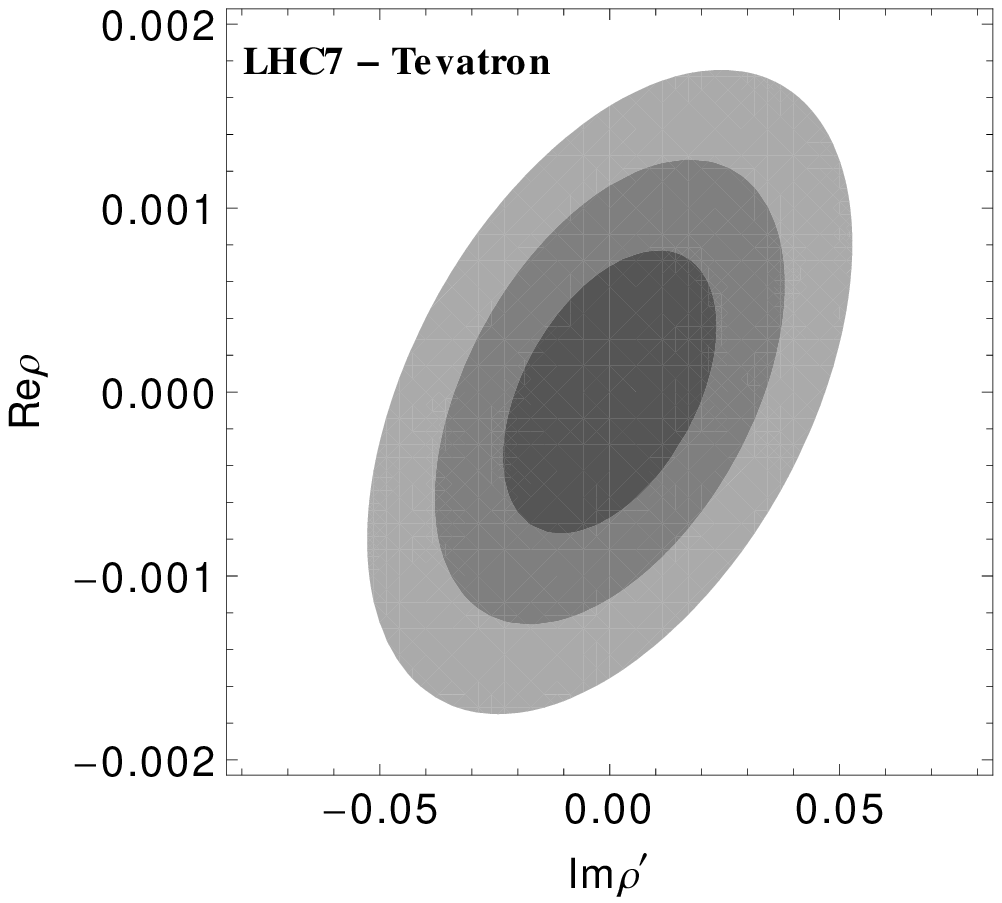} 
 \includegraphics[width=2.0in]{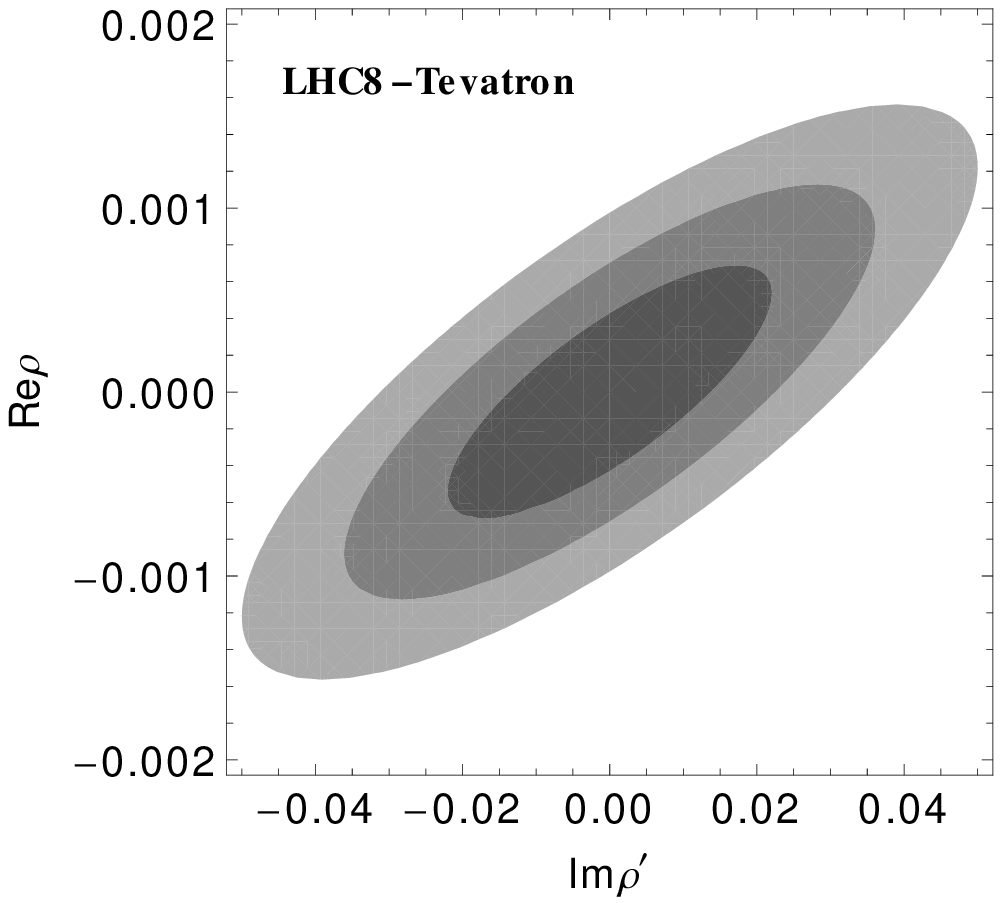} 
 \includegraphics[width=2.0in]{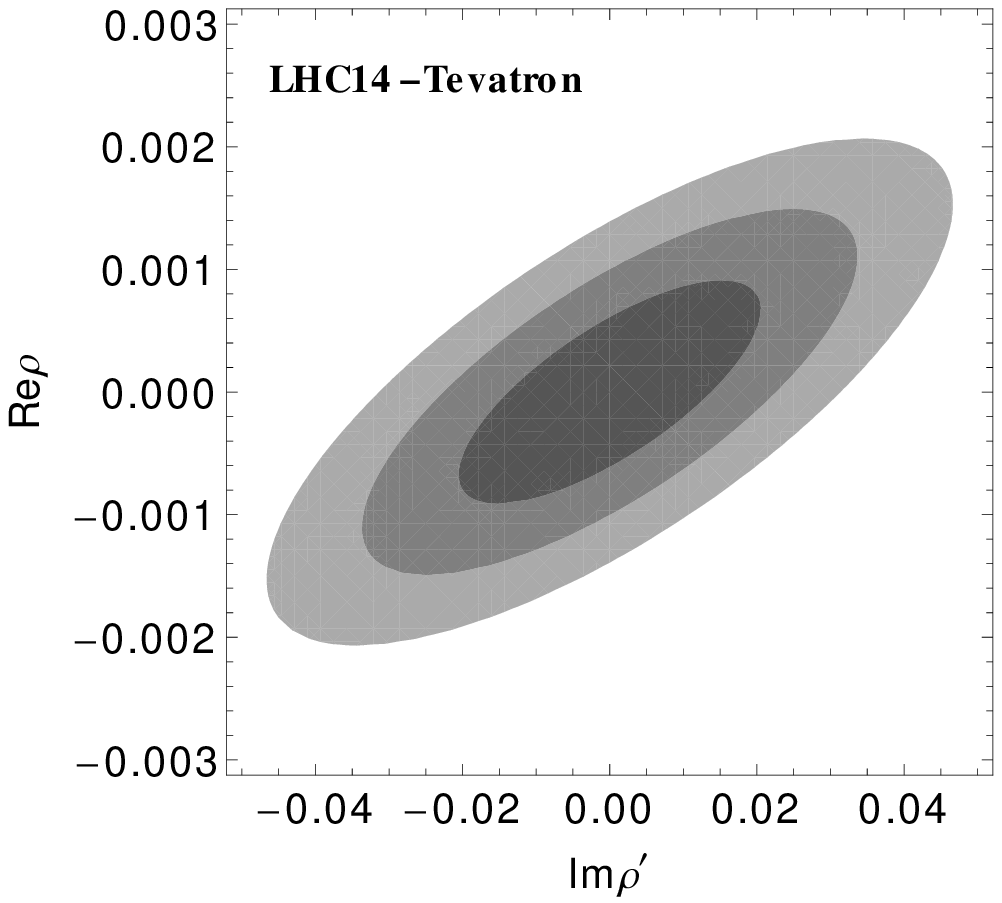} 
\caption{The 1$\sigma$ (central region), 2$\sigma$ (middle region) and 3$\sigma$ (outer region) CL regions in the 
Re$\rho$-Im$\rho^\prime$ plane allowed by the combined measurement of two observables at a time. The left, centre and right plots 
correspond to measurements at the combinations 
Tevatron-LHC7, Tevatron-LHC8 and Tevatron-LHC14 respectively. 
The $\chi^2$ values for 1$\sigma$, 2$\sigma$ and 3$\sigma$ CL intervals 
are 2.30, 6.18 and 11.83 respectively for 2 parameters in the fit.}
\label{lim-ind}
\end{center}
\end{figure}

For this, we perform a $\chi^2$ analysis to fit all the observables to within $f\sigma$ of statistical errors 
in the measurement of the observable. We define the following $\chi^2$ function 
\beq\label{chisq}
\chi^2 = \sum_{i=1}^n\left(\frac{P_i-O_i}{\sigma_i}\right)^2,
\eeq
where the sum runs over the $n$ observables measured and $f$ is the
degree of the confidence interval. $P_i$'s are the values of the observables
obtained by taking both anomalous couplings non-zero (and is a function of 
the couplings Re$\rho$ and Im$\rho^\prime$) and $O_i$'s are the values of the observables 
obtained in the SM. $\sigma_i$'s are the statistical fluctuations in the measurement of the observables, 
given in Eq. (\ref{stat-dev}).

In Fig. \ref{lim-ind}, we show the 1$\sigma$, 2$\sigma$ and 3$\sigma$ regions in Re$\rho$-Im$\rho^\prime$ 
plane allowed by combined measurement of asymmetry $A_\phi$ at different
experiments, taken two at a time. 
For this, in the $\chi^2$ function of Eq. (\ref{chisq}), we have combined the measurement at Tevatron with the measurements at
LHC7, LHC8 and LHC14.  From among the three combinations shown in Fig. \ref{lim-ind}, we find that the strongest simultaneous limits come from the 
combined measurements at Tevatron and LHC14, viz.,  $\pm0.006$  on Re$\rho$  and $\pm0.04$ on Im$\rho^\prime$,  at the $3\sigma$ level.

 We now describe some other relevant work on the determination of the chromomagnetic and 
chromoelectric form factors of the top at hadron colliders. 
Some earlier work \cite{atwood,haberl,hioki} made projections for
possible limits on the couplings which would be obtained at the
Tevatron and the LHC. 
In Ref. \cite{hioki}, the authors study polar-angle, 
transverse momentum and energy distributions of charged leptons coming
from top decay at the Tevatron and LHC. Our results 
on polar-angle distributions of charged leptons are in agreement with theirs
for nonzero $\Rerho2$.
They find 10-15\% deviations from the SM in the angular distributions for values of 
$\rho$ around 0.01 and of $\rho^\prime$ around 0.05. They also study
lepton energy distributions, which bring in dependence on anomalous
$tbW$ couplings. Refs. \cite{atwood,haberl} proposed utilizing cross section measurements at the LHC and 
the Tevatron to put limits on the anomalous couplings. With the
available data from Tevatron, Choudhury et
al.\cite{saha} using a slightly different notation, put limits on the new physics scale $\Lambda$.
They conclude that the cross section measurements at Tevatron would put a
lower bound 
on $\Lambda$ of about 7.4 TeV and 9 TeV for $\rho=\pm 1$ respectively  which in our notation would translate to
$\rho\sim[-1.94, 2.36]\times 10^{-2}$ while at LHC7 the lower bound on
$\Lambda$ is 10 TeV which is equivalent to  $\rho<1.75\times 10^{-2}$. 
Hioki and Ohkuma \cite{hioki2} in their work, which they consider an
update of \cite{haberl},
find that the Tevatron cross section
results  give bounds
$-.01<\rho < .01$ and $0.38<\rho < 0.41$ and
$|\rho^{\prime}|<0.12$, of which, only the region around $\rho=0$ and
$\rho^{\prime}=0$ survive on using early LHC data. They also studied the effect 
on top $p_T$, polar-angle distributions and invariant mass distributions from various combinations of 
$\rho$ and $\rho^\prime$ in the range of $0.1-0.4$ and found them to 
give significant deviations from SM predictions. 
Hesari and Najafabadi \cite{hesari}
studied the fraction of the $gg$ fusion contribution in $t\bar t$ production 
cross section at the Tevatron and at the LHC and concluded that this fraction 
is more sensitive at the Tevatron than at the LHC. From Tevatron results
with integrated luminosity of 1 fb$^{-1}$, they
quote limits  of $− 1.1 < \rho  < 0.6 , − 0.8 < \rho^{\prime} < 0.8.$
For the full luminosity accumulated at the Tevatron, they project limits
of
$−0.03 < \rho<
 1.5$ and $−0.37 < \rho^\prime< 0.37$ while 
at the LHC7, the limits they expect are $−0.04 < \rho < 0.98$ and $−0.15 <
\rho^\prime< 0.15$. They also studied the charge asymmetry of the top 
at the LHC and found very loose bounds on $\rho$ from it, and no
sensitivity to $\rho^\prime$. 

Other authors have considered possible limits on couplings from 
more detailed observations.
In Refs. \cite{gupta}, the authors considered probing
 the CP-violating chromoelectric 
dipole moment utilizing T-odd correlations constructed from jet and
lepton momenta and found the expected limit to be $|\rho^\prime|<5\times 10^{-3}$. 
Ref. \cite{Cheung:1995nt} studies the chromoelectric coupling utilizing various momentum correlations in $t\bar t$ and $t\bar t $ plus one-jet 
processes and derive the limit of $|\rho^\prime| \gtrsim 0.35$. In Ref.
\cite{Rizzo:1995uv} Rizzo studied anomalous $ttg$ couplings in single-top 
production at the Tevatron and at the LHC and concluded that the limits
from this channel are about one order of magnitude smaller than those from
the pair production 
processes. Effect of $\rho$ and $\rho^\prime$ on spin correlations in $t \bar t$ has been studied in Ref. \cite{Cheung:1996kc} 
and bounds are found to be $-0.7<\rho<0.6$ and $-0.5<\rho^\prime<0.5$.

In all above papers, authors have considered anomalous $ttg$ couplings to be real.
We consider both chromomagnetic and chromoelectric form factors to be
complex. We found, however, that our observables get contributions only
from the real part of $\rho$ and the imaginary part of $\rho^{\prime}$.
The projections for the best limits on $\Rerho2$ from our observables
are sometimes an order of magnitude better than those obtained from
cross sections.
In our analysis, we found top polarization and charge asymmetry 
of the charged lepton (both of which are CP odd) to be dependent on the imaginary part of the chromoelectric form factor. 
In Ref. \cite{Zhou:1998wz}, the author has considered chromomagnetic and chromoelectric form factors to be complex and construct various
CP violating observables to obtain constraints on real and imaginary
parts of $\rho^\prime$. At the Tevatron with 30 fb$^{-1}$, the author 
obtained limits of $2.4\times10^{-18}$ cm $g_s$ and $1.1\times 10^{-18}$ cm $g_s$ on Re$\rho^\prime$ and Im$\rho^\prime$ 
respectively which in our units translate to $2.13\times 10^{-2}$ and
$9.77\times 10^{-3}$. At LHC14 with 150 fb$^{-1}$, the limits obtained are 
$5.2\times10^{-20}$ cm $g_s$ and $2.5\times 10^{-20}$ cm $g_s$ on Re$\rho^\prime$ and Im$\rho^\prime$ 
respectively. In our units, limits are $4.62\times 10^{-3}$ and $1.91\times 10^{-3}$ on Re$\rho^\prime$ and Im$\rho^\prime$ 
respectively.
Considering that the luminosities we use for our limits are much lower,
our limits are comparable to theirs. 

More recently, Ref.~\cite{Englert:2012by}
has obtained upper bounds on chromomagnetic dipole moment
(Re$\rho$) as 0.085 using the available data for the cross sections for $t \bar t$ production
at Tevatron and LHC for 7 TeV. They have predicted that the sensitivity to
probe this coupling will be improved by a factor upto 4 by the boosted top
measurements at 14 TeV LHC. 


\section{Conclusions}
We have investigated the sensitivity of the Tevatron, LHC7, LHC8 and LHC14 to the anomalous $ttg$ couplings in top-pair production 
followed by semileptonic decay of the top. We derived analytical expressions for the spin density matrix for top-quark production
including the contributions of both real and imaginary parts of
anomalous $ttg$ couplings. We evaluate these at leading order in the strong
coupling, and neglect electroweak contributions. We work 
in the linear approximation of
anomalous couplings.
We find that only Re$\rho$ and Im$\rho^\prime$ give significant contributions to the spin density 
matrix at linear order. 
It may be noted that Im$\rho$ and Re$\rho^\prime$ do not appear in the
observables we consider. This may be understood from the fact that the
observables are even under naive time reversal T, viz., under change of
sign of all the momenta and spins, without an interchange of initial and
final states. In such a case, from the CPT theorem, the observables
which are CP even can only arise from dispersive parts of form factors
(in this case Re$\rho$) and the CP-odd observable can arise only from
absorptive parts (in this case Im$\rho^\prime$).

Longitudinal top polarization can be utilized to 
separate the contribution of Im$\rho^\prime$ completely independent of all other anomalous couplings whereas the total cross section
can be used to separate the contribution of Re$\rho$. 

Since top polarization can be measured only through the differential distribution of its decay products, we also 
study the angular distributions of the charged lepton coming from the
decay of the top. Charged-lepton momenta are  
measurable very accurately at the LHC and charged leptons have  the best spin analyzing power. 
Also, charged-lepton angular distributions have been shown 
to be independent of any NP in top decay. We find that the polar-angle distribution is 
not very sensitive to the anomalous couplings. On the other hand, the normalized azimuthal distribution is found to be sensitive to 
the anomalous couplings. The azimuthal distribution peaks
close to $\phi=0$ and $\phi = 2\pi$, and the values at the peaks are
quite sensitive to the magnitude and the sign of the anomalous couplings  
In order to quantify this difference and to be statistically more
sensitive, we construct an integrated azimuthal 
asymmetry from the azimuthal distribution of charged lepton. 

We study the effects of top transverse momentum cuts on top polarization and azimuthal distributions. We find that 
the top $p_T$ cut may enhance or reduce the top polarization depending whether we take a sample of low-$p_T$ or high-$p_T$ tops. 
In our analysis, we observed that for the Tevatron an ensemble of high-$p_T$ tops have higher degree of longitudinal top 
polarization as the function of imaginary part of anomalous chromoelectric coupling $\rho^\prime$. 
Conversely, an ensemble of low-$p_T$ tops reduce the top polarization at Tevatron for non-zero Im$\rho^\prime$. On the other hand, 
in the case of the LHC, the observation is reversed. For high-$p_T$ tops, top polarization is small and vice-versa.

We consider two angular asymmetries, leptonic charge asymmetry $A_{\rm ch}(\theta_0)$ and leptonic 
left-right asymmetry $A_\phi$ which serve as  measures of the anomalous couplings. 
We find that the $A_{\rm ch}(\theta_0)$ is proportional to $\imrho3$. 
The difference in the
$\ell^+$ and $\ell^-$ cross sections is relatively large in the range
$\theta_0:[\pi/8,3\pi/8]$. We choose the
cut-off $\theta_0$ to be $\pi/8$ to maximize the $A_{\rm ch}(\theta_0)$. We furthermore study the effects of 
top $p_T$ cuts on the charge asymmetry and conclude that $A_{\rm ch}(\theta_0)$ is enhanced in the 
low top-$p_T$ region at the LHC while the reverse is true for Tevatron.

The effect of top $p_T$ cuts on the angular distribution of charged leptons can be easily understood 
through Eq. (\ref{angdist}). For high-$p_T$ tops, the angular distributions peak at extreme 
values leading to larger azimuthal asymmetries $A_\phi$. We also study the effect of these cuts on the 
limits obtained by the measurement of top polarization and azimuthal asymmetry. 
We infer that the high-$p_T$ tops give large azimuthal asymmetries and can thus give more stringent limits on 
the chromomagnetic top-gluon coupling as compared to low-$p_T$ tops.  

We have restricted ourselves to an analysis of the statistical
sensitivities and not done a detailed analysis of the effects of cuts needed for
discrimination against background and of detector efficiencies. Such an
analysis would be required for a more precise determination of the
sensitivities of our observables. 
In conclusion, we have shown that top polarization, and subsequent
decay-lepton distributions can be used to
obtain fairly stringent limits on chromomagnetic and chromoelectric top
couplings from the existing Tevatron data as well as data soon to be
available after the 8 TeV run of the LHC. The limits could be improved
by the future runs of the LHC.

\begin{acknowledgements}
 SDR thanks the Theoretical Physics Department of Tata Institute of Fundamental Research for hospitality during the 
beginning of this work, and the Korea Institute for Advanced Study, Seoul for hospitality during the conclusion of 
the work. He also acknowledges financial support from the Department of Science and Technology, India, in the 
form of the J.C. Bose National Fellowship, grant no. SR/S2/JCB-42/2009.
We thank Namit Mahajan for pointing out an error in an earlier version
of the manuscript.
\end{acknowledgements}

\appendix
\section{Spin Density Matrix for top/anti-top in top-pair production with anomalous $ttg$ couplings}
In this Appendix, we present the spin density matrix elements
$\sigma^{ij}$ for the
top quark in the top pair production process. 
We include contributions of all anomalous couplings to linear order. 

Some general considerations can be used to anticipate the structure of
the density matrix. Writing the density matrix as a sum of
various contributions, $\sigma_{\rm SM}$ from the SM, and $\sigma_{   
\Rerho2,\rerho3,\Imrho2,\imrho3}$, the respective contributions from
$\Rerho2$, $\rerho3$, $\Imrho2$ and $\imrho3$,
\begin{equation}\label{sigma_anom}
\sigma^{ij} = \sigma^{ij}_{\rm SM} + \sigma^{ij}_{\Rerho2} +
\sigma^{ij}_{\rerho3} + \sigma^{ij}_{\Imrho2} + \sigma^{ij}_{\imrho3}.
\end{equation}
Then, Hermiticity of the density matrix gives
\begin{equation}\label{herm1}
\sigma^{\pm\pm} = \sigma^{\pm\pm *},
\end{equation}
implying that the diagonal matrix elements are real, and
\begin{equation}\label{herm2}
\begin{array}{ccc}
{\rm Re}~\sigma^{\pm\mp} &=& {\rm Re}~\sigma^{\mp\pm}\\ 
{\rm Im}~\sigma^{\pm\mp} &=& - {\rm Im}~\sigma^{\mp\pm}.
\end{array}
\end{equation}
Thus, the only imaginary contributions come in the off-diagonal
elements, changing sign under helicity flip.

Let us now see what transformation under naive time reversal T tells us.
Under T, $\sigma^{ij}$ is transformed to $\sigma^{ij*}$.
We note that $\rho^{\prime}$ terms are odd under CP, and therefore under
T. Hence terms with $\rerho3$ would change sign under T.
However, since T does not interchange initial and final states, the
the contribution from $\imrho3$ does not change sign, as
it arises from the absorptive part of some amplitude in
the underlying theory.
By the same argument, the $\Imrho2$ term changes sign under T, even though
the corresponding interaction is T invariant. 
We thus have the relations:
\begin{equation}\label{Ttransf}
\begin{array}{c}
\sigma^{\pm\pm}_{\rerho3} = \sigma^{\pm\pm}_{\Imrho2} = 0,
\\
{\rm Re}~\sigma^{\pm\mp}_{\rerho3}  = {\rm Re}~\sigma^{\pm\mp}_{\Imrho2} = 0,
\\  
{\rm Im}~\sigma^{\pm\mp}_{\rm SM} = {\rm Im}~\sigma^{\pm\mp}_{\Rerho2}  = 
{\rm Im}~\sigma^{\pm\mp}_{\imrho3} = 0.
\end{array}
\end{equation}
Thus, the diagonal density matrix elements can depend only on the
couplings $\Rerho2$
and $\imrho3$. 

The following relations arise from the parity transformation P, which
flips the signs of the helicities, using the fact
that the $\rho^{\prime}$ couplings are odd under P, and that there is an
extra phase factor of $-1$ in the transformation of the off-diagonal
elements:
\begin{equation}\label{Ptransf}
\begin{array}{c}
\sigma^{\pm\pm}_{\rm SM} = \sigma^{\mp\mp}_{\rm SM},\\
\sigma^{\pm\pm}_{\Rerho2} = \sigma^{\mp\mp}_{\Rerho2},\\
\sigma^{\pm\pm}_{\imrho3} = - \sigma^{\mp\mp}_{\imrho3},\\
{\rm Re}~\sigma^{\pm\mp}_{\rm SM} = -{\rm Re}~\sigma^{\mp\pm}_{\rm SM},\\
{\rm Re}~\sigma^{\pm\mp}_{\Rerho2} = -{\rm Re}~\sigma^{\mp\pm}_{\Rerho2},\\
{\rm Re}~\sigma^{\pm\mp}_{\imrho3} =  {\rm Re}~\sigma^{\mp\pm}_{\imrho3},\\
{\rm Im}~\sigma^{\pm\mp}_{\Imrho2} = - {\rm Im}~\sigma^{\mp\pm}_{\Imrho2},\\
{\rm Im}~\sigma^{\pm\mp}_{\rerho3} =  {\rm Im}~\sigma^{\mp\pm}_{\rerho3}.
\end{array}
\end{equation}
Now the above equations, together with the hermiticity relations 
in eq. \ref{herm2} tell us that 
\begin{equation}\label{Pandherm2}
\begin{array}{c}
{\rm Re}~\sigma^{\pm\mp}_{\rm SM}=0,\\
{\rm Re}~\sigma^{\pm\mp}_{\Rerho2}=0,\\
{\rm Im}~\sigma^{\pm\mp}_{\rerho3}=0.
\end{array}
\end{equation}

We conclude from the above that the diagonal
matrix elements, which are all real, can only get contributions from the
SM, and from $\Rerho2$ and $\imrho3$. The off-diagonal elements have no
contribution from the SM (since the SM amplitudes are real at tree
level), and get a real contribution from $\imrho3$ and an imaginary
contribution from $\Imrho2$. Also, the $\imrho3$ contribution in the
diagonal elements and the $\Imrho2$ contribution in the off-diagonal elements
change sign under helicity flip.

The
density matrix $\bar \sigma$ for the spin of the 
top anti-quark is obtained by changing the sign of the
Im$\rho$ and 
Im$\rho^\prime$ terms only in the off-diagonal element of the spin 
density matrix for the top quark. This can be seen from the fact that
under the operation of CPT, where T is naive time reversal (reversing
the sign of all spins and momenta, without interchange of initial and
final states), the top spin density matrix elements would be
transformed to the complex conjugates of the corresponding anti-top spin
density matrix elements, with the helicity indices
changing sign. However, 
this applies only to the real parts of
couplings. Contributions containing imaginary parts of
anomalous couplings, which arise from absorptive parts of amplitudes in
an underlying theory, would change sign under this operation. 
Thus, because of CPT invariance,
\begin{eqnarray}
\bar \sigma^{\pm\pm}_{\rm SM} = \sigma^{\mp\mp}_{\rm SM} =
\sigma^{\pm\pm}_{\rm SM},\\
\bar \sigma^{\pm\pm}_{\Rerho2} = \sigma^{\mp\mp}_{\Rerho2} =
\sigma^{\pm\pm}_{\Rerho2},\\
\bar \sigma^{\pm\pm}_{\imrho3} = - \sigma^{\mp\mp}_{\imrho3} =
\sigma^{\pm\pm}_{\imrho3},\\
{\rm Im}~\bar \sigma^{\pm\mp}_{\Imrho2} =  {\rm Im}~\sigma^{\mp\pm}_{\Imrho2} =
-{\rm Im}~\sigma^{\pm\mp}_{\Imrho2},\\
{\rm Re}~\bar \sigma^{\pm\mp}_{\imrho3} = -{\rm Re}~\sigma^{\mp\pm}_{\imrho3} =
-{\rm Re}~\sigma^{\pm\mp}_{\imrho3},
\end{eqnarray}

We denote the spin density matrix for the top quark in the $q\bar q$-initiated process as $\sigma^{\lambda\lambda^\prime}_{q\bar q}$ and
that for the gluon-gluon fusion process as $\sigma_{gg}^{\lambda\lambda^\prime}$. The labels (in subscript) $s$, $t$, $u$  
in $\sigma_{gg}^{\lambda\lambda^\prime}$ denote the s-, t- and u-channels contributions respectively whereas 
$st$, $su$ and $tu$ denote the interference between s- and t-channels, s- and u-channels and t- and u-channels 
respectively. 
The $\lambda$ and $\lambda^\prime$ are top helicities and may take values $\pm$. The total 
contribution of four-point $ggtt$ couplings is included in the terms
corresponding to the interference of the $s$-channel exchange amplitude
with the $t$- and $u$-channel exchange amplitudes with a coefficient 
labelled by $g^*$ and later set to 1.
The spin-density matrix elements $\sigma^{\lambda\lambda^\prime}$ for top quark in $t\bar t$-pair production 
in the parton cm frame (at parton level) are written as :
\begin{eqnarray}
\sigma^{++} &=& \MsqRR + \schRR + \tchRR + \uchRR + \stchRR + \suchRR + \tuchRR 
\\
\sigma^{+-} &=& \MsqRL + \schRL + \tchRL + \uchRL + \stchRL + \suchRL + \tuchRL 
\\
\sigma^{-+} &=& \MsqLR + \schLR + \tchLR + \uchLR + \stchLR + \suchLR + \tuchLR 
\\
\sigma^{--} &=& \MsqLL + \schLL + \tchLL + \uchLL + \stchLL + \suchLL + \tuchLL 
\end{eqnarray}
where 
\begin{eqnarray}
\MsqD &=& \label{qq-D}2\CqQs \hat{s}^2\left[16 \ \Rerho2 \pm 8 \ \imrho3 \ \bt\sstw +1+\bt^2\cstw+4r_t\right]\\
\MsqO &=& \frac{4\CqQs}{\Mt}\hat{s}^2 \ \sqrt{\hat{s}}\sin 2\theta_t \ 
\bt\left[\mp i \ \Imrho2 \ \bt + \imrho3\right]\\
\schD &=& 2\Cs2 \hat{s}^2\left[8 \ \Rerho2 \pm 8 \ \imrho3 \ \bt\cstw +(1-\bt^2\cstw)\right]\\
\schO &=&- \frac{4\Cs2}{\Mt}\hat{s}^2 \ \sqrt{\hat{s}}\sin 2\theta_t \ 
\bt\left[\pm i \ \Imrho2 \ \bt + \imrho3\right]
 \end{eqnarray}
 \begin{eqnarray}
\tchD &=& \nonumber \Ct2 \hat{s}^2 \left[16\Rerho2(1-\bt\ct3)\pm4\imrho3\left\{ \bt(1-8r_t)-\ct3(1+12r_t)\right.\right.\\\nonumber
&&+\bt\cstw(1+16r_t)-3\bt^2\cst+2\bt^3\csf\left.\left.\right\}+1+4r_t-16 r_t^2\right.\\
&&-\bt^3\ct3-8\bt^2r_t\cstw+\bt^3\cst-\bt^4\csf \left.\right]\\\nonumber
\tchO &=& \frac{2\Ct2}{\Mt}\hat{s}^2\sqrt{\hat{s}}\st3\left[\left(\imrho3 \pm i \bt \ \Imrho2\right)  
\left\{1+8r_t - 2\bt\ct3-8\bt\ct3 r_t \right.\right.\\
&& +\ 3 \bt^2\cstw - 2\bt^3\cst \left.\left.  \right\} 
+4r_t \ \imrho3\left\{1-\bt\ct3\right\}\right]\\
 \uchD &=& \nonumber \Cu2 \hat{s}^2 \left[16\Rerho2(1+\bt\ct3)\pm4\imrho3\left\{ \bt(1-8r_t)+\ct3(1+12r_t)\right.\right.\\\nonumber
 &&+\bt\cstw(1+16r_t)+3\bt^2\cst+2\bt^3\csf\left.\left.\right\}+1+4r_t-16r_t^2\right.\\
 &&+\bt^3\ct3-8\bt^2r_t\cstw-\bt^3\cst-\bt^4\csf \left.\right]\\\nonumber
%
\uchO &=& -\frac{2\Cu2}{\Mt}\hat{s}^2\sqrt{\hat{s}}\st3\left[\left(\imrho3 \pm i \bt \ \Imrho2\right)  
\left\{1+8r_t + 2\bt\ct3+8\bt\ct3 r_t \right.\right.\\
&& +\ 3 \bt^2\cstw + 2\bt^3\cst \left.\left.  \right\} 
+4r_t \ \imrho3\left\{1+\bt\ct3\right\}\right]\\
\stchD &=& 2\nonumber \CsCt \hat{s}^2\left[4 \ \Rerho2 (3-(2+g^*)\bt\ct3)\mp4 \ \imrho3\ct3 (2+g^*-2\bt^2\sstw-3\bt\ct3)\right.\\
&&+1-\bt^2\cstw-\bt^3\ct3\sstw \left. \right]\\
\stchO &=& \nonumber \frac{4 \CsCt}{\Mt}\ \hat{s}^2\sqrt{\hat{s}}\ \st3 \left[(\imrho3\pm \csqt\  \bt\Imrho2)(1-3\bt\ct3+2\bt^2\cstw+4r_t)\right.\\ 
&+&\left.4g^*r_t\imrho3\right]\\
\suchD &=& -2\nonumber \CsCu \hat{s}^2\left[4 \ \Rerho2 (3+(2+g^*)\bt\ct3)\pm4 \ \imrho3\ct3 (2+g^*-2\bt^2\sstw+3\bt\ct3)\right.\\
&&+1-\bt^2\cstw+\bt^3\ct3\sstw \left. \right]\\
\suchO &=&\nonumber  -\frac{4 \CsCu}{\Mt}\ \hat{s}^2\sqrt{\hat{s}}\ \st3 \left[(\imrho3\pm \csqt\  \bt\Imrho2)(1+3\bt\ct3+2\bt^2\cstw+4r_t)\right.\\ 
&+&\left.4g^*r_t\imrho3\right]\\
\tuchD &=& 2\CtCu \ \hat{s}^2 \ \sstw\bt\left[\pm 4 \ \imrho3 \left\{-1+2\bt^2\sstw\right\}
+\bt\{1-\bt^2\sstw\}\right]\\
\tuchO&=&-\frac{4\CtCu}{\Mt}\ \hat{s}^2 \ \sqrt{\hat{s}} \ \sin 2\theta_t \ \bt \left[ (\imrho3\pm \csqt \bt \ \Imrho2) \ \bt^2 \sstw  
- 2r_t\imrho3\}\right]
\end{eqnarray}

Here 
\begin{eqnarray} 
\CqQs & = & \frac{1 \ g_s^4}{18 \ \hat{s}^2};~~~~~~
\Cs2  =  \frac{3\ g_s^4}{64 \ \hat{s}^2};~~~~~~
\Ct2  =  \frac{g_s^4}{48 \ \left( \hat{t} - \Mt^2 \right)^2}\\
\Cu2 & = & \frac{g_s^4}{48 \ \left( \hat{u} - \Mt^2 \right)^2};~~~~~~~~
\CsCt  =  \frac{3\ g_s^4}{128 \ \hat{s} \ \left( \hat{t} - \Mt^2 \right)}\\
\CsCu & = & \frac{-3\ g_s^4}{128 \ \hat{s} \ \left( \hat{u} - \Mt^2 \right)};~~~~~~
\CtCu  =  \frac{-g_s^4}{384  \ \left( \hat{t} - \Mt^2 \right)\ \left( \hat{u} - \Mt^2 \right)}\\
r_t &=& \frac {\Mt^2}{\hat{s}};~~~~~~~~~~~~~~~~~~~~~~~~~~~
\bt = \sqrt{1-4\frac{\Mt^2}{\hat{s}}};~~~~~~~g^*=1.
\end{eqnarray}

\end{document}